\documentclass[a4paper,11pt]{article}
\usepackage{aaskaiid}
\usepackage{orcidlink}

\usepackage[normalem]{ulem} 

\newcommand{\soutPC}{\bgroup\markoverwith{\textcolor{cyan}{\rule[0.5ex]{2pt}{1pt}}}\ULon}

\newcommand{\soutrev}{\bgroup\markoverwith{\textcolor{red}{\rule[0.5ex]{2pt}{1pt}}}\ULon}

\newcommand{\soutrevnew}{\bgroup\markoverwith{\textcolor{cyan}{\rule[0.5ex]{2pt}{1pt}}}\ULon}
\defcitealias{Mangiagli_et_al_2022}{M22}
\defcitealias{Dong-Paez_et_al_2023b}{D23}

\title{Multi-messenger and Multi-band Studies of Massive Black Holes: the Synergies Between LISA and SKAO}
\ShortTitle{The Synergies Between LISA and SKAO}

\author[1]{Pedro~R. Capelo\orcidlink{0000-0002-1786-963X}}
\ShortName{P.~R. Capelo et al.}
\author[2]{Alberto Mangiagli\orcidlink{0000-0002-3689-1664}}
\author[3]{Lorenz Zwick\orcidlink{0000-0003-4818-3400}}
\author[1]{Lucio Mayer\orcidlink{0000-0002-7078-2074}}
\author[4]{Marta Volonteri\orcidlink{0000-0002-3216-1322}}

\affiliation[1]{Department of Astrophysics, University of Zurich, Winterthurerstrasse 190, CH-8057 Z\"urich, Switzerland}
\emailAdd{pcapelo@physik.uzh.ch}
\affiliation[2]{Max Planck Institute for Gravitational Physics (Albert Einstein Institute), Am M\"uhlenberg 1, DE-14476 Potsdam, Germany}
\affiliation[3]{Niels Bohr International Academy, Niels Bohr Institute, Blegdamsvej 17, DK-2100 K\o{}benhavn, Denmark}
\affiliation[4]{Institut d’Astrophysique de Paris, Sorbonne Universit\'{e}, CNRS, UMR 7095, 98 bis bd Arago, FR-75014 Paris, France}

\abstract{Depending on its mass, the same gas-embedded massive black hole (BH) binary can emit gravitational waves (GWs) in one or more bands (nHz and mHz) concurrently, along with electromagnetic (EM) waves over the entire spectrum. The Square Kilometre Array Observatory (SKAO), thanks to its unparalleled sensitivity, will be pivotal in achieving coincident GW-EM (mHz-radio) and GW-GW (nHz-mHz) detections. We review the state-of-the-art predictions -- achieved by means of numerical simulations and mock catalogues -- of the numbers of detectable coincident GW-EM signals when employing the Laser Interferometer Space Antenna (LISA) and SKA-Mid. By exploiting the same underlying BH binary populations, to allow for a fairer comparison, we then assess the importance of a variety of EM models for the radio flares and jets, finding that the number of radio counterparts of LISA sources is relatively insensitive to the jet/flare model employed and to whether SKA-Mid AA* or AA4 is assumed. Additionally, we describe how SKAO -- as part of a pulsar timing array (PTA) -- and LISA will provide the opportunity to detect the first low-frequency, multi-band (nHz-mHz) GW detection of the same object. Supermassive BH binaries embedded in gas discs are subjected to hydrodynamical torques, causing perturbations that produce additional small-amplitude, higher-frequency GWs. The main carrier GWs may thus be identifiable as a deterministic signal by SKAO-era PTAs, while the higher-frequency harmonics would shine in LISA as stochastic signals. Correlating these multi-band GWs would provide unprecedented constraints on the environment of the most massive BHs.
}

\begin{document}

\maketitle

\section{Introduction}

The final outcome of the merger of two massive galaxies hosting massive black holes (MBHs) is the emission of gravitational waves \citep[GWs;][]{Einstein_1916,Einstein_1918,Einstein_Rosen_1937}, produced by the changing quadrupole moment of the resulting MBH binary (MBHB), along with that of electromagnetic (EM) waves, originated by the gas surrounding the MBHB, both before and after the BH coalescence \citep[e.g.][]{DeRosa_et_al_2019,Bogdanovic_et_al_2022,Krause_et_al_2025}.

For an MBHB mass of $\sim$$10^4$--$10^7$~M$_{\odot}$, the GW signal will be detected in the $\sim$mHz range by the Laser Interferometer Space Antenna \citep[LISA;][]{Amaro-Seoane_et_al_2023,Colpi_et_al_2024} and other future space-based detectors (e.g. TianQin; \citealt{Luo_et_al_2016,Li_et_al_2025}; and Taiji; \citealt{Ruan_et_al_2020,Luo_et_al_2021}). More massive MBHBs, with masses in the range $\sim$$10^8$--$10^{10}$~M$_{\odot}$, instead emit GWs in the $\sim$nHz range and are thus detectable by pulsar timing arrays \citep[PTAs; e.g.][]{Agazie_et_al_2023_Constraints}, although current PTAs are sensitive only to the background generated by many systems, rather than by individual MBHBs \citep[e.g.][]{Agazie_et_al_2023_Bayesian}.

The same mergers that originate the MBHB can also generate strong gas inflows from tidal torques, due to both gravitational \citep[e.g.][]{Barnes_Hernquist_1991,Hopkins_Quataert_2010,Capelo_et_al_2015} and hydrodynamical \citep[e.g.][]{Barnes_2002,Capelo_Dotti_2017,Blumenthal_Barnes_2018} forces which transport gas from galactic scales down to the influence radius of the MBHB and below. There, the gas can be found in the form of a rotating circumnuclear disc at hundreds of parsec scales \citep[][]{Mayer_2013}, as suggested by observations of galaxy nuclei \citep[][]{Medling_et_al_2014}, with further outward transport of angular momentum resulting in a circumbinary disc or an accretion disc, depending on the mass ratio of the MBHBs \citep[][]{SouzaLima_et_al_2020,2021dan}. In a circumbinary disc, the MBHBs are fed by gas streams that penetrate the cavity and generate mini-discs around each of the two MBHs, the significance of which depends on gas thermodynamics \citep[][]{Krause_et_al_2025}. These disc-like structures can all produce EM emission in various forms \citep[see, e.g.][]{Bogdanovic_et_al_2022}. In particular, the EM emission can take place before the emission of GWs as well as being concurrent with the latter, or occur post-merger as an afterglow. The ensuing EM waves cover the entire spectrum, from X-ray to radio, the latter related to diffuse radiation as well as jets produced or modulated during the BH coalescence, and detectable by the Square Kilometre Array Observatory \citep[SKAO;][]{Carilli_Rawlings_2004,bourke2015advancing} in the frequency range 50--15400~MHz. It is noteworthy to recall the the only detected MBHs with a separation small enough, of order 7~pc, to be an actually bound binary of MBHs has been discovered in the radio, appearing as a source with a double radio core \citep[][]{Rodriguez_et_al_2006}.

Moreover, the gas surrounding these MBHBs may also have an effect on the dynamics of the binary. Under certain conditions, the presence of stochastic gas torques induces the emission of GW harmonics, which shine at frequencies higher than the main GW emission \citep[][]{Zwick_et_al_2022}. In the case of a very massive binary, whose signal may be detected by a future SKAO-like PTA, these so-called dirty waveforms (DWs) could then be detectable in the mHz range, yielding the very first low-frequency multi-band GW detection of the same system.\\

The coincident detection of these different kinds of radiation (GWs, EM waves, and DWs) will provide invaluable information on the compact objects themselves and on the binary's environment. Since LISA and SKAO will operate at the same time (in the late 2030s), this will give us a tremendous opportunity to learn more about the evolution of MBHs and their hosts.

In Section~\ref{sec:EM-mHz}, we review recent studies on joint GW-radio detections of MBHBs of mass $\sim$$10^4$--$10^7$~M$_{\odot}$ and, by adopting the same underlying MBH population, we re-compute the numbers of events, to obtain a more proper comparison.

In Section~\ref{sec:nHz-mHz}, we describe how an SKAO-era PTA may be able to detect individual MBHBs of mass $\sim$$10^8$--$10^{10}$~M$_{\odot}$ in the nHz range and discuss the possibility of detecting higher-frequency waves (in the mHz range) originating from the same system, caused by the gas surrounding the binary.

\section{Radio counterparts of LISA massive black hole binaries}\label{sec:EM-mHz}

In order to predict the number of joint GW-radio multi-messenger detections, we require a model for the underlying MBHB population one adopts, as well as on the assumed EM model and on the selected observational constraints. Here we compare two recent state-of-the-art works, those of \citeauthor{Mangiagli_et_al_2022} (\citeyear{Mangiagli_et_al_2022}; hereafter \citetalias{Mangiagli_et_al_2022}) and \citeauthor{Dong-Paez_et_al_2023b} (\citeyear{Dong-Paez_et_al_2023b}; hereafter \citetalias{Dong-Paez_et_al_2023b}), both building on the works of \citet{Barausse_2012} and \citet{Tamanini_et_al_2016}, in order to understand the origin of their different outcomes and inform future studies, by highlighting where the (radio) community should focus their modelling efforts.

\citetalias{Mangiagli_et_al_2022} used the results of the \textsc{BACH} \citep[Black holes Across Cosmic History;][]{Barausse_2012,Sesana_et_al_2014,Antonini_et_al_2015a,Antonini_et_al_2015b} semi-analytical model to track the evolution of the relevant properties of the merging MBHBs and their host galaxies (specifically, BH masses and spins, and the surrounding gas). From those, they computed the EM detectability of each MBHB, assuming several observatories and wavelength ranges. Here, we focus on the radio regime and in particular on SKAO, for which they modelled the flare emission,\footnote{A flare can be defined as any brightening in the radio band not associated to the jet(s). In this work, we focus on one specific type of flare, associated to the twisting of magnetic field lines following the motion of the MBHB, but note that future investigations may eventually uncover other contributions to the radio detectability of a binary.} following \citeauthor{Palenzuela_et_al_2010} (\citeyear{Palenzuela_et_al_2010}; see also \citealt{Kaplan_et_al_2011,Neilsen_et_al_2011,OShaughnessy_et_al_2011}), and the jet luminosity, after \citet{Meier_et_al_2001}; considered different collimations for the flare and the jet; and finally imposed a radio flux threshold of $F_{\rm lim} = \nu F_{\nu,\rm lim}$, where $\nu = 1.7$~GHz and $F_{\rm \nu,lim} = 1\, \mu$Jy was the SKA-Mid spectral flux density limit assumed at that frequency. For the subset of radio-detected systems, \citetalias{Mangiagli_et_al_2022} then computed the GW signal \citep[adopting the inspiral-merger-ringdown waveform \textsc{PhenomHM};][]{London_et_al_2018}, imposed a GW signal-to-noise ratio (SNR) $> 10$, and required that the systems satisfy an SKAO sky localization threshold of $\Delta \Omega < 10$~deg$^2$, finally obtaining the number of joint GW-radio detections:\footnote{\label{foot:ELT}\citetalias{Mangiagli_et_al_2022} additionally imposed that the redshift of the source should be measured independently, assuming that the extremely large telescope (ELT) can be used. For this work, we do not require this additional constraint. We note, however, that the results do not change significantly (compare their table~V with their table~VIII), because the chosen sky localization requirement selects relatively low-redshift systems, which can mostly be detected by ELT.} the average number of multi-messenger systems assuming 4~yr of LISA observations (i.e. 5~yr of observations with an 80 per cent duty cycle), regardless of the kind of BH seed and the assumed delay between the galaxy merger and the MBHB coalescence, is of the order of a few tens, when considering an isotropic flare and a jet collimation with a Lorentz factor of $\Gamma = 2$ ($\theta = \Gamma^{-1} \simeq 30^{\circ}$; `maximizing scenario'), of the order of a few, when considering a collimation of $\Gamma = 2$ for both the flare and the jet (`minimizing scenario'), and close to zero, when the collimation for both phenomena is $\Gamma = 10$ ($\theta \simeq 6^{\circ}$; see their table~VIII).

\citetalias{Dong-Paez_et_al_2023b} adopted the results of the \textsc{Obelisk} cosmological, radiation-hydrodynamical simulation of a protocluster progenitor \citep[run down to redshift $z = 3.5$;][]{Trebitsch_et_al_2021,Dong-Paez_et_al_2023a}, from which they retrieved MBHB and host quantities. For the jet luminosity, they either (`pessimistic scenario') used the so-called fundamental plane of BH activity \citep[][]{Gultekin_et_al_2009}, which links the (core) radio luminosity at 5~GHz, the integrated X-ray luminosity in the 2--10~keV range, and the BH mass, or (`optimistic scenario') adopted a jet-formation model following \citet{Meier_et_al_2001} -- although with slightly different parameters than in \citetalias{Mangiagli_et_al_2022} (see below) -- and then calculated the luminosity at 2~GHz for both models assuming a power-law spectrum $L_{\nu} \propto \nu^{-0.7}$. For the flare luminosity, they assumed an increase of the pre-merger jet luminosities dependent on the mass ratio. They then imposed an SKA-Mid radio spectral flux density threshold of 4.2~$\mu$Jy, computed assuming 9~hr exposure observations at 2~GHz (and that the observed sources are not resolved), and a full-SKA-Mid limit of 0.4~$\mu$Jy. They finally computed the joint GW-radio detections using the same methodology and constraints of \citetalias{Mangiagli_et_al_2022}. However, since \textsc{Obelisk} simulates a highly biased region of the Universe, thus not representative of the cosmic population as a whole, computing the event rates is not straightforward (see \citealt{Izquierdo-Villalba_et_al_2026} for an attempt).

Indeed, interpreting the differences between \citetalias{Mangiagli_et_al_2022} and \citetalias{Dong-Paez_et_al_2023b} is made difficult by the fact that they used not only different EM models and spectral flux density limits, but also distinct underlying MBHB populations. \citetalias{Mangiagli_et_al_2022} used an MBHB population derived from a semi-analytical galaxy formation and evolution model, whereas \citetalias{Dong-Paez_et_al_2023b} employed a catalogue built from a hydrodynamical simulation (of a biased region of the Universe, run down to $z = 3.5$). This resulted, amongst other things, in different distributions of MBHB mass ratios, therefore affecting the results, since the mass ratio is a crucial indicator of how much gas reaches the remnant's central region \citep[see, e.g.][]{Capelo_et_al_2017} and thus directly affects the EM emission \citep[see, e.g.][for the flare luminosity]{Kaplan_et_al_2011}. In particular, the population used by \citetalias{Dong-Paez_et_al_2023b} has a larger fraction of low-mass-ratio systems than that employed by \citetalias{Mangiagli_et_al_2022} (compare \citetalias{Dong-Paez_et_al_2023b}'s figure 9 with \citetalias{Mangiagli_et_al_2022}'s figure 16). In this chapter, however, we are more interested in understanding the role of the EM models, rather than finding the rather complex dependence on the MBHB population. For this reason, we decided to use the same MBHB populations for both models, selecting that used in \citetalias{Mangiagli_et_al_2022}, which we briefly describe here.

The MBHB populations are constructed based on three different models \citep[see][]{Barausse_2012,Klein_et_al_2016,Tamanini_et_al_2016}: a light-seed prescription (`PopIII'), yielding Pop~III seeds of mass $\sim$$10^2$--$10^3$~M$_{\odot}$ at $z \sim 15$--20, which accounts for the delays between galaxy mergers and BH coalescences; and a heavy-seed recipe, producing heavy seeds of mass $\sim$$10^4$--$10^5$~M$_{\odot}$ at $z \sim 8$--15, which accounts (`Q3d'; where `Q3' stands for the \citealt{Toomre_1964} parameter being equal to 3 in the \textsc{BACH} model and `d' stands for delay) or does not account (`Q3nd'; where `nd' stands for no delay) for the above mentioned delays (see \citealt{Antonini_et_al_2015b,Tamanini_et_al_2016} for details). The MBHB population is evolved from $z=20$ to $z=0$. Using catalogues simulating 90 yr of data, the total numbers of MBHBs for the models PopIII, Q3d, and Q3nd are 15546, 692, and 10700, respectively. A summary of the MBHB models adopted in this study can be found in Table~\ref{tab:summary_SAM_models}.

\begin{table}
\centering
\caption{Summary of the astrophysical MBH models adopted in the text. The merger rates were computed by dividing the total number of MBHBs by the simulated time of the catalogues (90~yr).}
\begin{tabular}{c|c|c|c|c}
\hline
Model & Mass seed range [M$_{\odot}$] & Initial redshift range & Delays & Merger rate [$\rm yr^{-1}$] \\
\hline  
PopIII  & $\sim$$10^2$--$10^3$  & $\sim$15--20  & Yes   & 172.7 \\
Q3d     & $\sim$$10^4$--$10^5$  & $\sim$8--15   & Yes   & 7.7   \\
Q3nd    & $\sim$$10^4$--$10^5$  & $\sim$8--15   & No    & 118.9 \\
\hline
\end{tabular}
\vspace{-0.1cm}
\label{tab:summary_SAM_models}
\end{table}

For clarity, we write here the EM models tested in this work, highlighting also the differences with past work. We include the production of jets from accreting spinning BHs, due to the \citet{Blandford_Znajek_1977} effect, and that of flares, originating from the twisting of magnetic field lines by the inspiralling BHs \citep[e.g.][]{Palenzuela_et_al_2010}.

The jet luminosity is computed in three different ways, the first two accounting for the total luminosity of the jet and the third only taking into account the contribution of the core radio luminosity:

\begin{itemize}

\item {\it Jet-M01/M22} -- The first method, adopted by \citet{Tamanini_et_al_2016} and \citetalias{Mangiagli_et_al_2022}, follows the recipe by \citet{Meier_et_al_2001}, who provided the total jet power from radiatively efficient, optically thick, geometrically thin, cold accretion discs \citep[][]{Shakura_Sunyaev_1973} and radiatively inefficient, optically thin, geometrically thick, hot, advection-dominated accretion flows \citep[ADAFs;][]{Narayan_and_Yi_1994}:\\
\begin{equation}
\frac{L_{\rm jet,M01}}{{\rm erg~s}^{-1}} = \begin{cases}
    10^{42.7} \eta_{\rm S} \left(\frac{\alpha}{0.01}\right)^{-0.1} m_9^{0.9} \left(\frac{\dot{m}_{\rm jet}}{0.1}\right)^{6/5} (1+1.1a_1+0.29a_1^2) {\rm ~for~geom.~thin~discs}\,,\\
    10^{45.1} \eta_{\rm S} \left(\frac{\alpha}{0.3}\right)^{-1} m_9 \left(\frac{\dot{m}_{\rm jet}}{0.1}\right) g^2 (0.55f^2 + 1.5 f a_1 +a_1^2) {\rm ~for~geom.~thick~discs}\,,
    \label{eq:jetMeier}
\end{cases}
\end{equation}

\noindent where the viscosity parameter $\alpha = 0.1$, $m_9$ is the mass of the primary BH in units of $10^9$~M$_{\odot}$, $\dot{m}_{\rm jet}$ is the BH accretion rate in units of the \citet{Eddington_1916} limit (when assuming a radiative efficiency $\epsilon_{\rm r} = 0.1$; see below), $a_1$ is the spin magnitude of the primary BH, and $f = 1$ and $g = 2.3$ regulate the angular velocity of the gas and the azimuthal magnetic field of the system, respectively \citep[][]{Meier_et_al_2001}. \citetalias{Mangiagli_et_al_2022} employed the formula for the geometrically thin discs when the \citeauthor{Eddington_1916} ratio $f_{\rm Edd} \le 0.3$ and that for the geometrically thick discs when $f_{\rm Edd} > 0.3$ (thereby assuming that ADAFs and slim discs have similar jet powers), imposing a floor at $f_{\rm Edd} = 0.02$ and a ceiling at $f_{\rm Edd} = 1$. In this work, we define $f_{\rm Edd} = L_{\rm bol}/L_{\rm Edd}$, with $L_{\rm bol}$ being the bolometric luminosity of the binary (computed as $\epsilon_{\rm r} \dot{M}_{\rm acc}c^2$, with $\epsilon_{\rm r} = 0.1$, $c$ the speed of light in vacuum, and $\dot{M}_{\rm acc}$ the accretion rate onto the binary, calculated as the amount of gas surrounding the binary at merger divided by the viscous time-scale; see \citetalias{Mangiagli_et_al_2022} for more details) and $L_{\rm Edd}$ its \citeauthor{Eddington_1916} luminosity, defined as $4 \pi G (m_1+m_2) m_{\rm p}c/\sigma_{\rm T}$, with $G$, $m_{\rm p}$, and $\sigma_{\rm T}$ being the gravitational constant, the proton mass, and the \citet{Thomson_1906} scattering cross-section, respectively, and $m_1$ and $m_2$ the masses of the primary and secondary BH, respectively. The jet power is then assumed to be all emitted at 1.4~GHz (corresponding to the nominal frequency of SKA-Mid Band~2, which is currently the highest-frequency band below $\sim$5~GHz to be deployed), i.e. $\eta_{\rm S} = 1$, and beamed with an opening angle $\theta = 1/\Gamma$, where $\Gamma = 2$ is the Lorentz factor \citep[][]{Cohen_et_al_2007}. The latter effect introduces  two opposing effects on jet observability: one the one hand, if the line of sight between the observer and the source is outside the opening angle, then no radio emission can be detected; on the other hand, if the observer is inside the opening angle, they will observe an increased flux due to collimation.

\item {\it Jet-M01/D23} -- The second method, adopted by \citetalias{Dong-Paez_et_al_2023b}, also follows the recipe by \citeauthor{Meier_et_al_2001} (\citeyear{Meier_et_al_2001}; Equation~\ref{eq:jetMeier}), but with different parameters and assumptions,\footnote{\label{foot:DP}We note that the normalizations employed by \citetalias{Dong-Paez_et_al_2023b} were also different, as they used $10^{43.5}$ and $10^{46}$ for the geometrically thin and thick discs, respectively. Additionally, they did not apply any beaming to the jet emission. However, for a fairer comparison with other models, we use here the published values by \citet{Meier_et_al_2001} and collimate the jet as in the {\it Jet-M01/M22} model.} as \citetalias{Dong-Paez_et_al_2023b} (i) assumed different thresholds for the disc regimes, employing the formula for the geometrically thin discs when $f_{\rm Edd} \ge 10^{-2}$ and that for the geometrically thick discs (in this case, ADAFs) when $f_{\rm Edd} < 10^{-2}$, imposing no floor and no ceiling for $f_{\rm Edd}$, (ii) employed $\alpha = 0.1$ for the geometrically-thin-disc case and $\alpha = 0.3$ for the geometrically-thick-disc case, (iii) set $\eta_{\rm S} = 10^{-2}$ (to account for the fact that not all jet power is emitted at 1.4~GHz), and (iv) assumed no beaming (although in this work we apply beaming; see Footnote~\ref{foot:DP}).

\item {\it Core-FP/G19} -- The third method, adopted by \citetalias{Dong-Paez_et_al_2023b}, assumes that the rest-frame radio and X-ray luminosities and the BH mass follow the so-called fundamental plane (FP) of BH activity, and that this radio luminosity only takes into account the contribution of the core. Whereas \citetalias{Dong-Paez_et_al_2023b} adopted the relation by \citet{Gultekin_et_al_2009}, we here adopt the more recent relation by \citeauthor{Gultekin_et_al_2019} (\citeyear{Gultekin_et_al_2019}; although we note that the results adopting the relation by \citealt{Gultekin_et_al_2009} are almost indistinguishable):\\
\begin{equation}
    \log_{10} \left(\frac{L_{\rm R,FP/G19}}{10^{38}~{\rm erg~s}^{-1}}\right) = -0.62 + 0.70 \log_{10} \left(\frac{m_1+m_2}{10^8~{\rm M}_{\odot}}\right) + 0.74 \log_{10} \left(\frac{L_{\rm X}}{10^{40}~{\rm erg~s}^{-1}}\right)\,,
\end{equation}

\noindent where $L_{\rm R}$ and $L_{\rm X}$ are the $\nu = 5$~GHz radio and 2--10~keV X-ray rest-frame luminosity, respectively, and we note that we used the total binary mass in the relation, even though the original formula assumed single BHs.\footnote{We also note that the FP of BH activity is known to hold for $f_{\rm Edd} \lesssim 10^{-2}$ \citep[e.g.][]{Gallo_et_al_2012}. However, we here assume that it holds for any value of $f_{\rm Edd}$ (as done by \citetalias{Dong-Paez_et_al_2023b}). We additionally caution that not all sources follow the FP, including BHs with masses in the LISA range \citep[see, e.g.][]{Gultekin_et_al_2022}.} The X-ray luminosity was computed from the bolometric luminosity (not limited by any $f_{\rm Edd}$ floor or ceiling), adopting the bolometric correction by \citet{Shen_et_al_2020}. We then compute the core luminosity at 1.4~GHz, assuming a power-law spectrum, $L_{\nu} \propto \nu^{- \alpha_{\rm r}}$, where $\alpha_{\rm r} = 0.7$ \citep[][]{Gultekin_et_al_2014}, thus obtaining $L_{\rm core,FP/G19}$. Since this is core emission, we did not apply any beaming.

\end{itemize}

\begin{figure}
\includegraphics[width=1\textwidth]{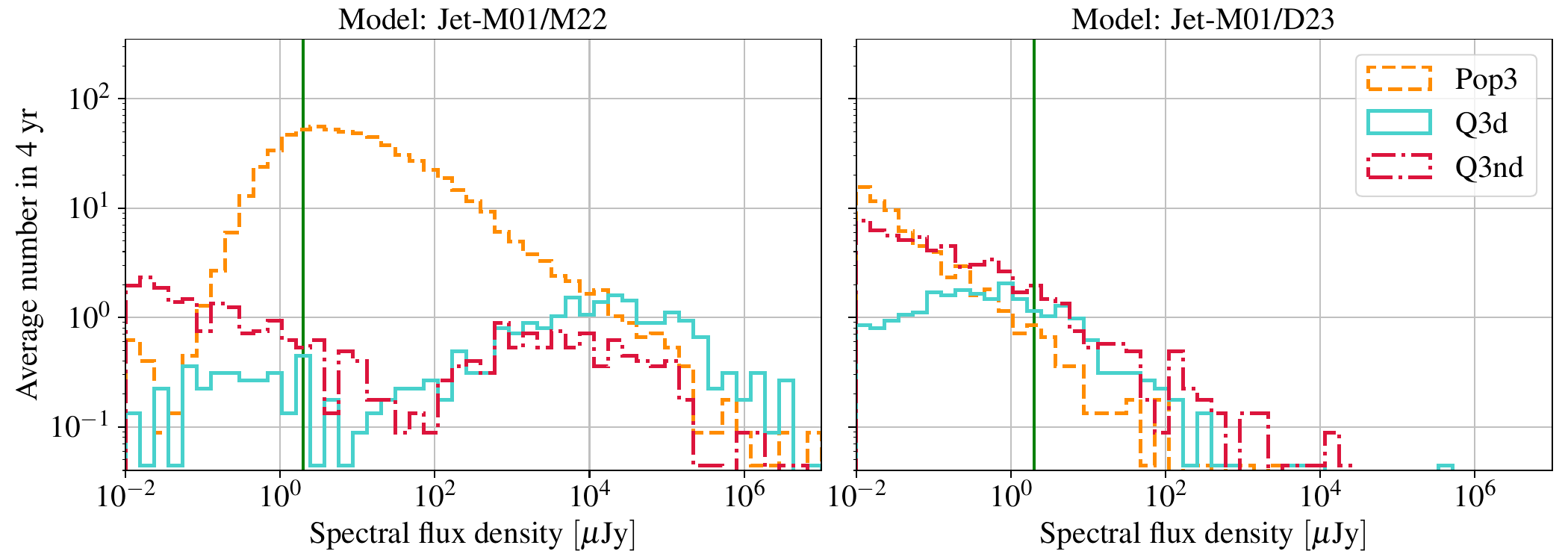}
\caption{Distribution of jet radio spectral flux densities according to two of the models discussed in the text. Left-hand panel: {\it Jet-M01/M22} -- \citet{Meier_et_al_2001} model for the total jet emission with the parameters from \citetalias{Mangiagli_et_al_2022}. Right-hand panel: {\it Jet-M01/D23} -- \citet{Meier_et_al_2001} model for the total jet emission with the (corrected; see Footnote~\ref{foot:DP}) parameters from \citetalias{Dong-Paez_et_al_2023b}. Different colours and linestyles correspond to three different populations of MBHBs as reported in the legend. The vertical line shows the assumed spectral flux density limit of SKA-Mid (array assembly 4: AA4) at 2~$\mu$Jy (for $\sim$4~hours of observing time in Band~2; see text for details). The integrals of these curves (and of those of Figures~\ref{fig:magn_ska_distro_cp_janski_onlyflare} and \ref{fig:magn_ska_distro_cp_janski}) over the total range of spectral flux densities are equal to four times the merger rates quoted in Table~\ref{tab:summary_SAM_models}.
}
\label{fig:magn_ska_distro_cp_janski_onlyjet}
\end{figure}

We compute the jet radio spectral flux densities -- assuming that the radio emission is isotropic (i.e. the ratio between luminosity and flux is given by $4 \pi d_{\rm L}^2$, where $d_{\rm L}$ is the luminosity distance) in the {\it Core-FP/G19} case and collimated (with $\Gamma = 2$) in the {\it Jet-M01/M22} and {\it Jet-M01/D23} cases \citep[see also the discussion in][]{Tamanini_et_al_2016} -- according to the three models described above, for the three different MBHB populations. In the left-hand and right-hand panels of Figure~\ref{fig:magn_ska_distro_cp_janski_onlyjet}, we show the spectral flux densities produced with the \citet{Meier_et_al_2001} jet model adopted by \citetalias{Mangiagli_et_al_2022} and \citetalias{Dong-Paez_et_al_2023b} (corrected; see Footnote~\ref{foot:DP}), respectively. The spectral flux densities in the latter case are much lower than in the former, mostly due to the relatively low value of $\eta_{\rm S}$ in their model. We do not show the spectral flux densities produced by the FP model, as they are much lower than the other two. However, as we will see, these differences become unimportant when we add a flare model.

The flare luminosity is computed in two different ways:

\begin{itemize}

\item {\it Flare-M22} -- The first model, adopted amongst others by \citetalias{Mangiagli_et_al_2022}, follows the work of \citet{Palenzuela_et_al_2010,Kaplan_et_al_2011,Neilsen_et_al_2011,OShaughnessy_et_al_2011}, yielding
\begin{equation}
    L_{\rm flare,M22} = f_{\rm Edd} f_{\rm R} \left(\frac{v}{v_{\rm max}}\right)^2 q^{-2} L_{\rm Edd}\,,
    \label{eq:flareM22}
\end{equation}

\noindent where $q = m_1/m_2$ is the mass ratio of the two BHs (with $m_1 \ge m_2$), $f_{\rm R}$ is the radio bolometric correction, and $v/v_{\rm max}$ describes the luminosity evolution as the binary inspirals. We set $f_{\rm R} = 0.1$ and $v/v_{\rm max} = 1$, as done by \citet{Tamanini_et_al_2016} and \citetalias{Mangiagli_et_al_2022}, and impose $0.02 \le f_{\rm Edd} \le 1$, as in the case of the {\it Jet-M01/M22} model.

\item {\it Flare-D23} -- The second method, adopted by \citetalias{Dong-Paez_et_al_2023b}, assumes that the flare luminosity is related to the jet luminosities as
\begin{equation}
    L_{\rm flare,D23} = (1 + \kappa q^{-2})(L_{\rm S,s1} + L_{\rm S,s2})\,,
    \label{eq:flareD23}
\end{equation}

\noindent where $\kappa$ parametrizes the increase in luminosity due to the flare, $L_{\rm S,s1}$ and $L_{\rm S,s2}$ are the jet luminosities from \citet{Meier_et_al_2001}, as implemented with the model \emph{Jet-M01/D23} (thus with no floor nor ceiling for $f_{\rm Edd}$), and we set $\kappa = 5$ (as done by \citetalias{Dong-Paez_et_al_2023b}).

\end{itemize}

\begin{figure}
\includegraphics[width=1\textwidth]{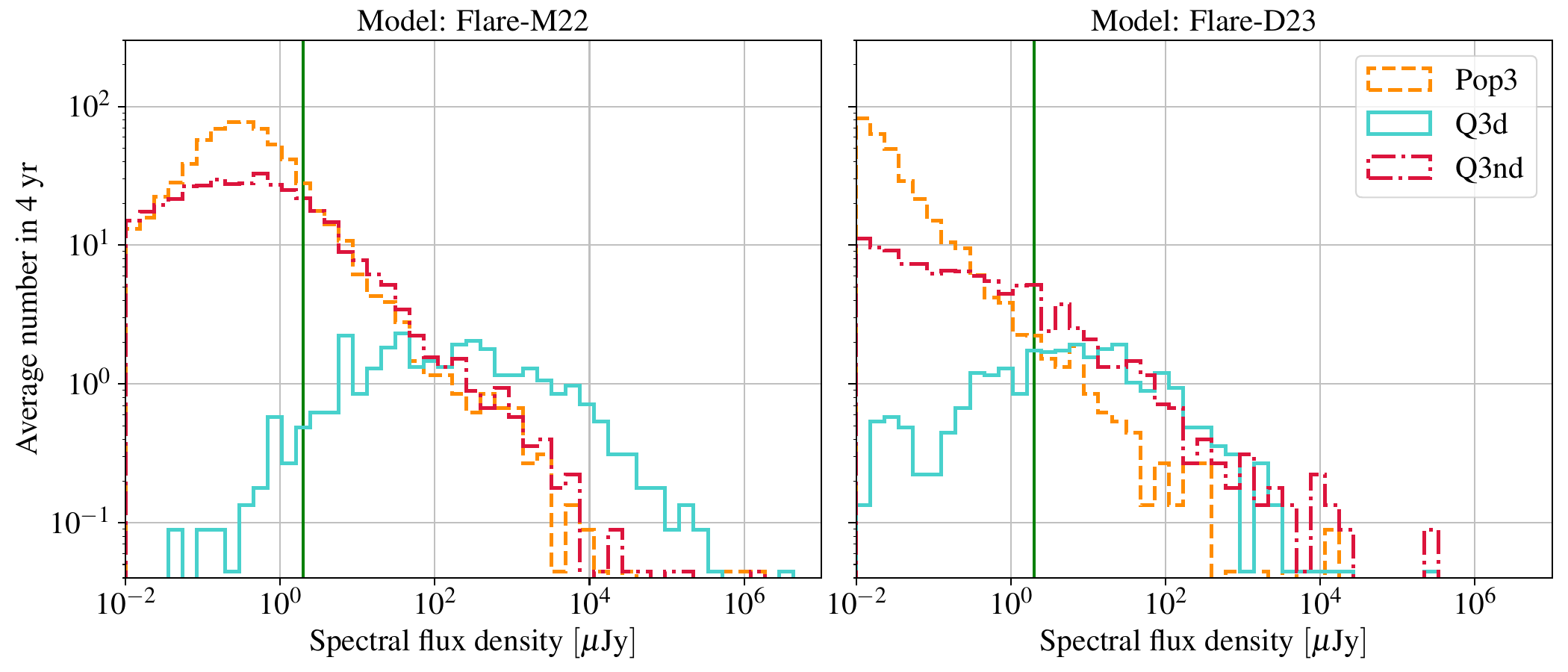}
\caption{Distribution of flare radio spectral flux densities according to the two models discussed in the text. Left-hand panel: {\it Flare-M22} -- adopted by \citetalias{Mangiagli_et_al_2022}. Right-hand panel: {\it Flare-D23} -- adopted by \citetalias{Dong-Paez_et_al_2023b}. Different colours and linestyles correspond to three different population of MBHBs as reported in the legend. The vertical line shows the assumed spectral flux density limit of SKA-Mid (AA4) at 2~$\mu$Jy.
}
\vspace{-0.4cm}
\label{fig:magn_ska_distro_cp_janski_onlyflare}
\end{figure}

In Figure~\ref{fig:magn_ska_distro_cp_janski_onlyflare}, we present the flare radio spectral flux densities -- computed assuming isotropy -- according to the two recipes just described, for the same MBHB populations assumed for Figure~\ref{fig:magn_ska_distro_cp_janski_onlyjet}. The two flare models produce similar spectral flux density distributions. Most importantly, we note that, for a given spectral flux density threshold (e.g. $\sim$1~$\mu$Jy), the number of sources with a detectable flare radio flux is larger than the number of systems with a detectable jet radio flux, also when considering the models for the total jet luminosity: this is because, even though (jet) collimation increases the flux received from the observer, thus allowing for the detection of farther and fainter systems, these additional sources do not make up for the lost systems which are not detected because the beam does not intersect the observer's line of sight, when compared to the number of detected sources from an isotropic flare.

We now consider four different scenarios, in which we combine the jet and flare contributions in different ways, to assess the relevance of each choice:

\begin{itemize}

\item {\it Scenario~1 -- Jet-M01/M22 + Flare-M22}: jet model by \citet{Meier_et_al_2001}, using the parameters of \citetalias{Mangiagli_et_al_2022}, and flare model by \citetalias{Mangiagli_et_al_2022}.

\item {\it Scenario~2 -- Core-FP/G19 + Flare-M22}: core model according to the FP by \citet{Gultekin_et_al_2019}, and flare model by \citetalias{Mangiagli_et_al_2022}.

\item {\it Scenario~3 -- Jet-M01/M22 + Flare-D23}: jet model by \citet{Meier_et_al_2001}, using the parameters of \citetalias{Mangiagli_et_al_2022}, and flare model by \citetalias{Dong-Paez_et_al_2023b}.

\item {\it Scenario~4 -- Core-FP/G19 + Flare-D23}: core model according to the FP by \citet{Gultekin_et_al_2019}, and flare model by \citetalias{Dong-Paez_et_al_2023b}.

\end{itemize}

\begin{figure}
\includegraphics[width=1\textwidth]{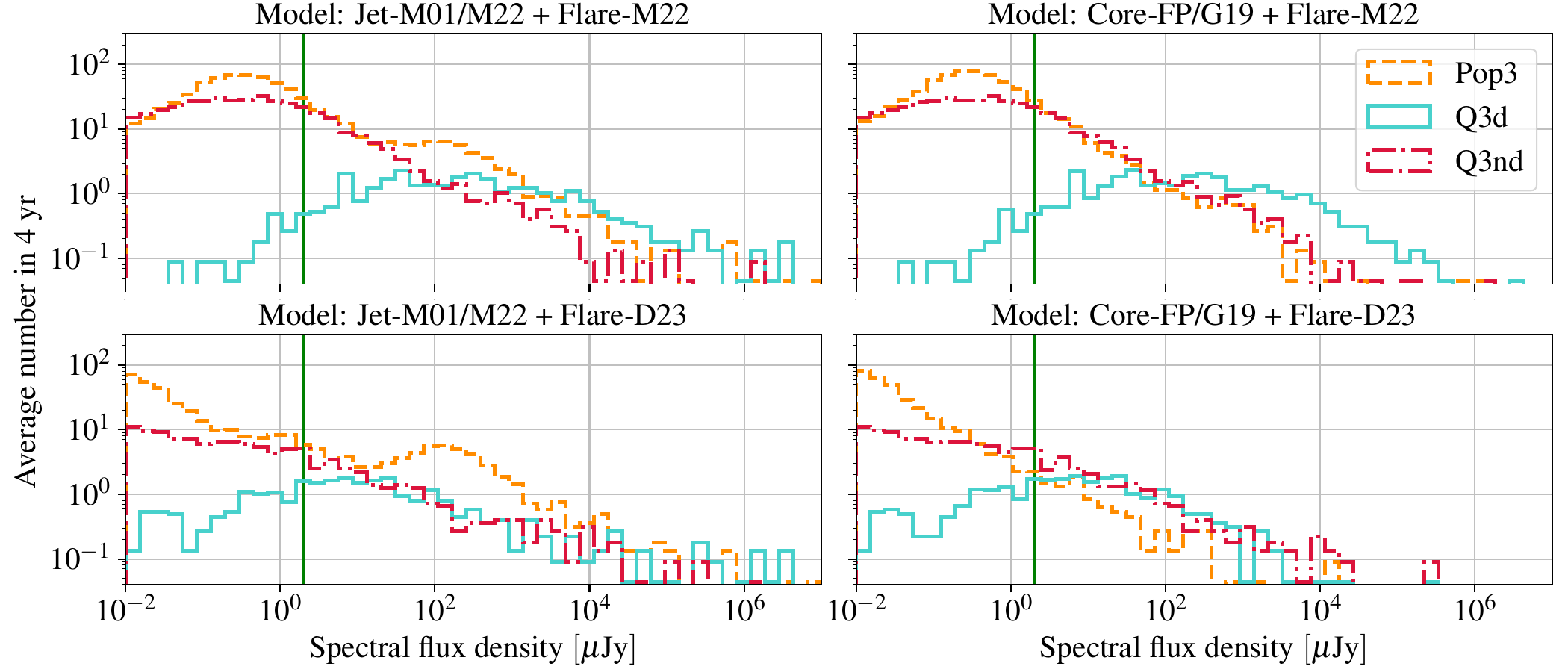}
\caption{Distribution of radio spectral flux densities according to the models discussed in the main text. Left-upper panel: {\it Jet-M01/M22 + Flare-M22}. Right-upper panel: {\it Core-FP/G19 + Flare-M22}. Left-lower panel: {\it Jet-M01/M22 + Flare-D23}. Right-lower panel: {\it Core-FP/G19 + Flare-D23}. Different colours and linestyles correspond to three different population of MBHBs as reported in the legend. The vertical line shows the assumed spectral flux density limit of SKA-Mid (AA4) at 2~$\mu$Jy. 
}
\label{fig:magn_ska_distro_cp_janski}
\end{figure}

\begin{table}
\centering
\caption{Average number of events with total spectral flux densities $>$2~$\mu$Jy (AA4 threshold; see text for details) and $>$2.7~$\mu$Jy (AA* threshold) in 4~yr from Figure~\ref{fig:magn_ska_distro_cp_janski}.}
\begin{tabular}{c|c|c|c|c}
\hline
& \begin{tabular}{@{}c@{}} Scenario 1: \\ Jet-M01/M22 \\ + Flare-M22 \\ \hline AA4 \;\;\; AA* \end{tabular} & \begin{tabular}{@{}c@{}} Scenario 2: \\ Core-FP/G19 \\ + Flare-M22 \\ \hline AA4 \;\;\; AA* \end{tabular} & \begin{tabular}{@{}c@{}} Scenario 3: \\ Jet-M01/M22 \\ + Flare-D23 \\ \hline AA4 \;\;\; AA* \end{tabular} & \begin{tabular}{@{}c@{}} Scenario 4: \\ Core-FP/G19 \\ + Flare-D23 \\ \hline AA4 \;\;\; AA* \end{tabular} \\
\hline  
PopIII  & 126.9 \;\;\; 109.4 & 79.6 \;\;\; 63.1 & 62.0 \;\;\; 57.9 & 9.2 \;\;\; 8.0 \\
Q3d     & 28.9 \;\;\; 28.7   & 28.8 \;\;\; 28.6 & 18.8 \;\;\; 17.6 & 18.0 \;\;\; 16.7 \\
Q3nd    & 84.6 \;\;\; 71.3   & 84.3 \;\;\; 70.9 & 23.2 \;\;\; 20.0 & 22.1 \;\;\; 18.9 \\
\hline
\end{tabular}
\label{tab:total_new}
\end{table}

The choice of these combinations was made in order to assess how many coincident radio-GW sources one could detect when assuming a very optimistic scenario, wherein both the jet and flare are maximized ({\it Scenario~1}), and a very pessimistic scenario, wherein both the jet and flare are minimized ({\it Scenario~4}), with two additional scenarios in between. The distributions of spectral flux densities of these four models are shown in Figure~\ref{fig:magn_ska_distro_cp_janski}. The distributions are overall quite similar (especially above the detection thresholds for SKAO, see below), mostly because of the similarity of spectral flux density distributions from the two models of the flare, which, being luminous and isotropic, produces more detectable systems.

\begin{figure}
\includegraphics[width=1\textwidth]{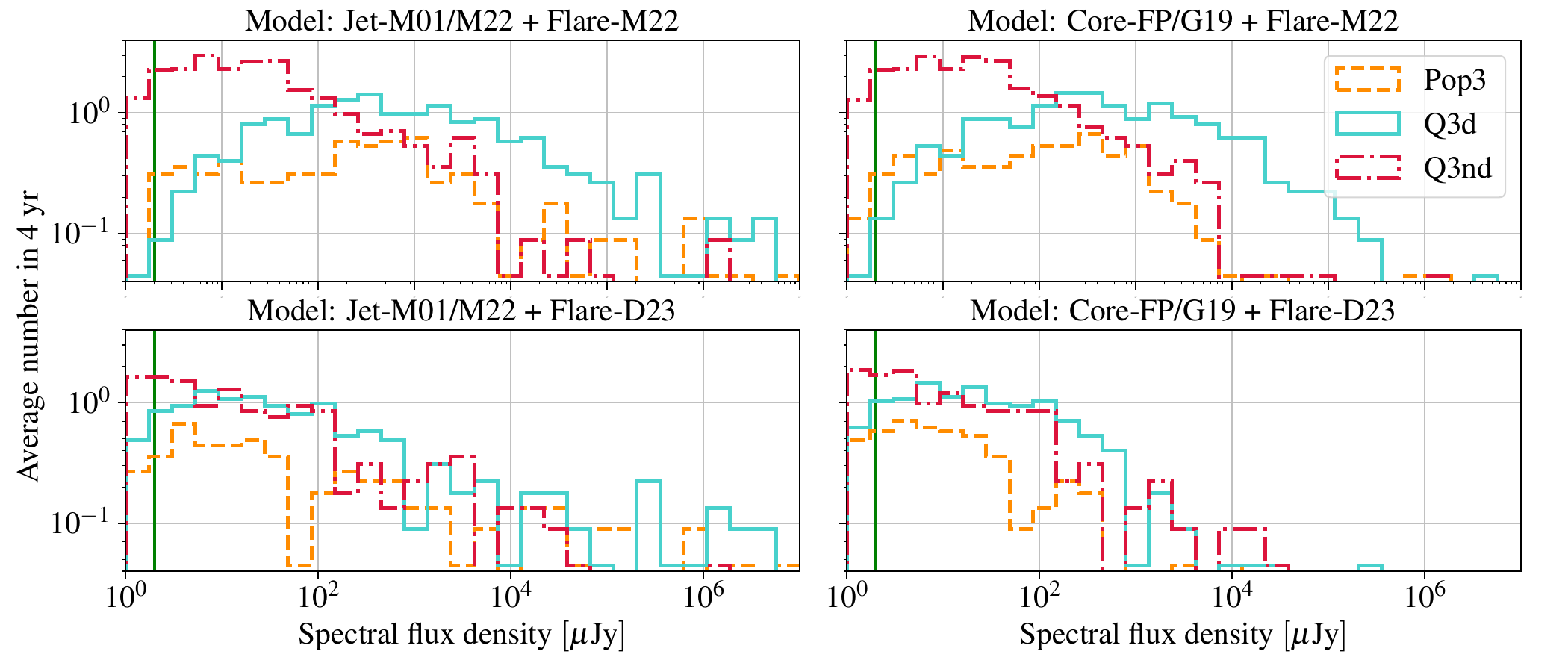}
\caption{Spectral flux density distributions for the EM counterparts we can detect with SKA-Mid for the four different scenarios discussed in the text. The vertical line shows the assumed spectral flux density limit of SKA-Mid (AA4) at 2~$\mu$Jy. 
}
\label{fig:magn_ska_distro_stsi_janski}
\end{figure}

\begin{table}
\centering
\caption{Average number of EM counterparts in 4~yr from Figure~\ref{fig:magn_ska_distro_stsi_janski}, assuming spectral flux density thresholds of 2~$\mu$Jy (SKA-Mid AA4) and 2.7~$\mu$Jy (SKA-Mid AA*).
}
\begin{tabular}{c|c|c|c|c}
\hline
& \begin{tabular}{@{}c@{}} Scenario 1: \\ Jet-M01/M22 \\ + Flare-M22 \\ \hline AA4 \;\;\; AA* \end{tabular} & \begin{tabular}{@{}c@{}} Scenario 2: \\ Core-FP/G19 \\ + Flare-M22 \\ \hline AA4 \;\;\; AA* \end{tabular} & \begin{tabular}{@{}c@{}} Scenario 3: \\ Jet-M01/M22 \\ + Flare-D23 \\ \hline AA4 \;\;\; AA* \end{tabular} & \begin{tabular}{@{}c@{}} Scenario 4: \\ Core-FP/G19 \\ + Flare-D23 \\ \hline AA4 \;\;\; AA* \end{tabular} \\
\hline  
PopIII  & 6.4 \;\;\; 6.0   & 6.1 \;\;\; 5.6   & 4.8 \;\;\; 4.3   & 4.2 \;\;\; 3.5 \\
Q3d     & 15.3 \;\;\; 15.2 & 15.2 \;\;\; 15.1 & 11.7 \;\;\; 10.6 & 11.2 \;\;\; 9.9 \\
Q3nd    & 22.9 \;\;\; 20.0 & 22.8 \;\;\; 19.9 & 10.8 \;\;\; 9.0  & 10.4 \;\;\; 8.5 \\
\hline
\end{tabular}
\vspace{0.5cm}
\label{tab:total_cp}
\end{table}

After computing the radio spectral flux densities (from flares and jets combined), we check if the source can be detected by the radio observatory, requiring that the radio flux $F_{\rm R} \ge F_{\rm min}$, where $F_{\rm min}$ is the detector's flux limit. The flux limit can be written as $F_{\rm min} = \nu F_{\rm \nu,min}$, where $\nu$ is the frequency at which the emission takes place and $F_{\rm \nu,min}$ is the detector's spectral flux density limit at such frequency, so that we can write\\
\begin{equation}
    F_{\rm R} \ge 10^{-20} \frac{\nu}{{\rm GHz}} \frac{F_{\rm \nu,min}}{{\rm \mu Jy}} {\rm ~erg~s}^{-1} {\rm cm}^{-2}\,.
\end{equation}

To compute such limit in the case of SKAO, we employ the radiometer equation, which provides the 1-$\sigma$ spectral flux density noise level:\\
\begin{equation}
    F_{\rm \nu,1\text{-}\sigma} = \frac{2 k_{\rm B} T_{\rm sys}}{\eta_{\rm s} A_{\rm eff} \sqrt{\Delta\nu \, t_{\rm obs}}} = \frac{2 k_{\rm B} T_{\rm sys}}{\eta_{\rm s} \eta_{\rm eff}N\pi(D/2)^2 \sqrt{\Delta\nu \, t_{\rm obs}}}\,,
\end{equation}

\noindent where $k_{\rm B}$ is the Boltzmann constant, $\eta_{\rm s}$ is the system efficiency, $T_{\rm sys}$ is temperature of the dishes, $A_{\rm eff}$ is their effective (combined) area, with $\eta_{\rm eff}$, $N$, and $D$ being the aperture efficiency, number, and diametre of the dishes, respectively, $\Delta\nu$ is the total bandwidth, $t_{\rm obs}$ is the observing time, and we assumed only one polarization channel.

We consider $\nu = 1.4$~GHz, i.e. the nominal frequency of SKA-Mid's Band~2, for which we assume $\eta_{\rm s} = 0.969$, $T_{\rm sys} = 13$~K, $\Delta\nu = 808$~MHz, $\eta_{\rm eff} = 0.79$, and $D = 15$~m, with the planned number of dishes for the SKA-Mid AA* and AA4 arrays being 144 and 197, respectively.\footnote{See the SKA1 Design Baseline Description document (SKA-TEL-SKO-0001075, Revision~02, Date 2022-01-24) at \url{https://zenodo.org/records/16895574}.
} With these numbers, assuming $\sim$4~hours of observing time and imposing a 5-$\sigma$ detection threshold (i.e. $F_{\rm \nu,min} = 5F_{\rm \nu,1\text{-}\sigma}$), we obtain $F_{\rm \nu,min} = 2$~$\mu$Jy and $= 2.7$~$\mu$Jy for the SKA-Mid AA4 and AA* array, respectively.

\begin{table}
\centering
\caption{Average number of EM counterparts in 4~yr when the jet luminosity is modelled with the model {\it Jet-M01/D23}, assuming spectral flux density thresholds of 2~$\mu$Jy (SKA-Mid AA4) and 2.7~$\mu$Jy (SKA-Mid AA*).
}
\begin{tabular}{c|c|c}
\hline
& \begin{tabular}{@{}c@{}} Jet-M01/D23 + Flare-M22 \\ \hline AA4 \;\;\; AA* \end{tabular} & \begin{tabular}{@{}c@{}} Jet-M01/D23 + Flare-D23 \\ \hline AA4 \;\;\; AA* \end{tabular}  \\
\hline  
PopIII  & 6.2 \;\;\; 5.7   & 4.3 \;\;\; 3.6 \\
Q3d     & 15.2 \;\;\; 15.1 & 11.4 \;\;\; 9.9 \\
Q3nd    & 23.0 \;\;\; 20.0 & 10.7 \;\;\; 8.7 \\
\hline
\end{tabular}
\label{tab:total_cp_JetD23}
\end{table}

We impose these two limits on the distributions shown in Figure~\ref{fig:magn_ska_distro_cp_janski}, obtaining the numbers given in Table~\ref{tab:total_new}. As expected, the most optimistic ({\it Scenario 1: Jet-M01/M22 + Flare-M22}) and most pessimistic ({\it Scenario 4: Core-FP/G19 + Flare-D23}) scenario give, respectively, the highest and lowest number of events, for all three populations, with the two mixed models yielding intermediate numbers. We also note that the rates do not vary significantly when changing the SKA-Mid configuration (from AA* to AA4), because the change in sensitivity given by the different number of dishes is not very large and also does not occur near any break in the spectral flux density distributions.

In order to compute the number of EM counterparts, we further impose a GW SNR > 10 and a sky localization $\Delta \Omega < 10$~deg$^2$ (as in \citetalias{Mangiagli_et_al_2022}), obtaining the distributions shown in Figure~\ref{fig:magn_ska_distro_stsi_janski} and the numbers of detections given in Table~\ref{tab:total_cp}. Here, the differences encountered in the number of events (see Figure~\ref{fig:magn_ska_distro_cp_janski} and Table~\ref{tab:total_new}) are even smaller, with the most optimistic combination generating only a factor of $\sim$2 more EM counterparts than the most pessimistic one. This is because the additional constraints naturally select systems at low redshift, which are then more easily detectable in either scenario. The natural follow-up of this computation would be determining the likelihood of finding these EM counterparts amongst other radio sources. This would require establishing the feasibility of identifying host galaxies at high redshift, which necessitates dedicated galaxy catalogues, a sophisticated model for the EM counterparts from other galaxies within a given error volume, and a statistical approach to deal with EM observations. One first step would be to extract the information on the luminosity distance of the event from the GW signal. This information, together with the sky localization, would allow us to construct an error volume in the sky and to exclude all the already-known sources that do not fall inside this error volume. This is beyond the scope of this chapter, but some preliminary works can be found in the literature \citep[e.g.][]{Ravi_2018,DalCanton_et_al_2019,Lops_et_al_2023}.

\begin{figure}
\includegraphics[width=1.0\textwidth,]{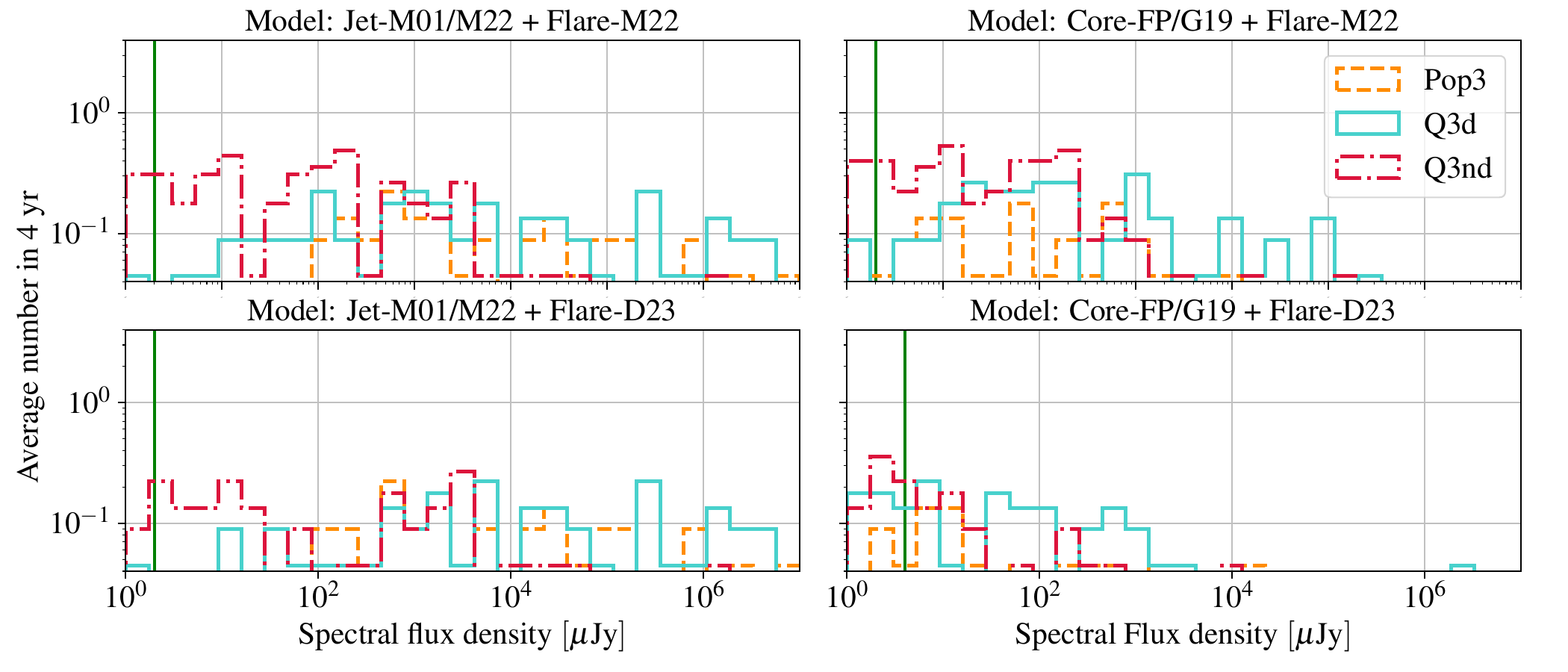}
\caption{Same as Figure~\ref{fig:magn_ska_distro_stsi_janski}, but with both the jet and flare collimated with an opening angle of $\sim$30$^\circ$ for the four scenarios presented in the text. The vertical line shows the assumed spectral flux density limit of SKA-Mid (AA4) at 2~$\mu$Jy. 
}
\label{fig:magn_ska_distro_stsi_janski_without_elt_collimatedflare}
\end{figure}

\begin{table}
\centering
\caption{Average number of EM counterparts in 4~yr from Figure~\ref{fig:magn_ska_distro_stsi_janski_without_elt_collimatedflare}, where both jet and flare are collimated, assuming spectral flux density thresholds of 2 (SKA-Mid AA4) and 2.7~$\mu$Jy (SKA-Mid AA*). We note that these are the same scenarios as in Tables~\ref{tab:total_new}--\ref{tab:total_cp}, with the only difference that the flare is collimated.
}
\begin{tabular}{c|c|c|c|c}
\hline
& \begin{tabular}{@{}c@{}} Jet-M01/M22 \\ + Flare-M22 \\ \hline AA4 \;\;\; AA* \end{tabular} & \begin{tabular}{@{}c@{}} Core-FP/G19 \\ + Flare-M22 \\ \hline AA4 \;\;\; AA* \end{tabular} & \begin{tabular}{@{}c@{}} Jet-M01/M22 \\ + Flare-D23 \\ \hline AA4 \;\;\; AA* \end{tabular} & \begin{tabular}{@{}c@{}} Core-FP/G19 \\ + Flare-D23 \\ \hline AA4 \;\;\; AA* \end{tabular} \\
\hline  
PopIII  & 1.6 \;\;\; 1.6 & 1.2 \;\;\; 1.1 & 1.5 \;\;\; 1.5 & 0.7 \;\;\; 0.6 \\
Q3d     & 2.8 \;\;\; 2.7 & 2.7 \;\;\; 2.6 & 2.1 \;\;\; 2.1 & 1.5 \;\;\; 1.2 \\
Q3nd    & 3.9 \;\;\; 3.5 & 3.7 \;\;\; 3.2 & 1.9 \;\;\; 1.7 & 1.2 \;\;\; 0.7 \\
\hline
\end{tabular}
\label{tab:total_cp_coll}
\end{table}

Reminiscent of the fact that the two models {\it Jet-M01/M22} and {\it Jet-M01/D23} produce very different jet radio spectral flux density distributions (compare the left-hand and right-hand panels of Figure~\ref{fig:magn_ska_distro_cp_janski_onlyjet}), we also considered two additional combined models, {\it Jet-M01/D23 + Flare-M22} and {\it Jet-M01/D23 + Flare-M22}, in which we simply replaced the jet model {\it Jet-M01/M22} with {\it Jet-M01/D23}, and re-computed the number of EM counterparts, shown in Table~\ref{tab:total_cp_JetD23}. Because of the dominant effect of the flare, which yields similar results between the models {\it Flare-M22} and {\it Flare-D23}, the numbers are very similar (compare the first and second column of Table~\ref{tab:total_cp_JetD23} with the first and second column of Table~\ref{tab:total_cp}, and the third and fourth column of Table~\ref{tab:total_cp_JetD23} with the fifth and sixth column of Table~\ref{tab:total_cp}).

Finally, we considered the case in which both the flare and the jet are collimated (both with $\Gamma = 2$, i.e. $\theta \simeq 30^{\circ}$), obtaining the distributions and numbers of events shown in Figure~\ref{fig:magn_ska_distro_stsi_janski_without_elt_collimatedflare} and Table~\ref{tab:total_cp_coll}, respectively. Overall, all the EM counterparts rates plummet, since a collimated flare, though having an increased flux and in principle favouring the detection of fainter and farther sources, in most cases has a beam that does not intersect the observer's line of sight.

In conclusion, by using the same underlying population of BH binaries, we were able to assess the effect of EM modelling. Using different combinations of three jet models and two flare models -- updated versions of those employed by \citetalias{Mangiagli_et_al_2022} and \citetalias{Dong-Paez_et_al_2023b} -- we computed the number of EM counterparts for each scenario, finding it to be quite independent of the EM choices we made, except of course for when we collimate the radio flare, and of the SKAO configuration. We caution, however, that this result is mostly a consequence of our two flare models yielding similar luminosities. We thus stress that the radio community should focus their efforts in building a reliable prediction for the flare emission.

\section{Multi-band gravitational-wave detections of gas-embedded massive black holes}\label{sec:nHz-mHz}

\subsection{Single-source detection of massive black holes with an SKAO-era pulsar timing array}

The direct detection of individual MBHBs (of mass $\sim$$10^8$--$10^{10}$~M$_{\odot}$) through their GW emission remains beyond the reach of current PTAs \citep[][]{2023arzu,2023falxa,Agazie_et_al_2023_Bayesian}, which are only marginally sensitive to the stochastic GW background generated by the superposition of many such binaries \citep[e.g.][]{Agazie_et_al_2023_Evidence,Agazie_et_al_2023_Constraints}. However, the situation is expected to improve dramatically with the advent of SKAO \citep[e.g.][]{Janssen_et_al_2015,Shannon_et_al_2025,Shannon01.2026.SKA}, thanks to the greatly enhanced ability to track numerous pulsars with improved timing precision, thereby increasing the chances of detecting individual binary systems \citep[][]{2009smits,2025depta}.

\begin{figure}
    \centering
    \includegraphics[width=0.666666\linewidth]{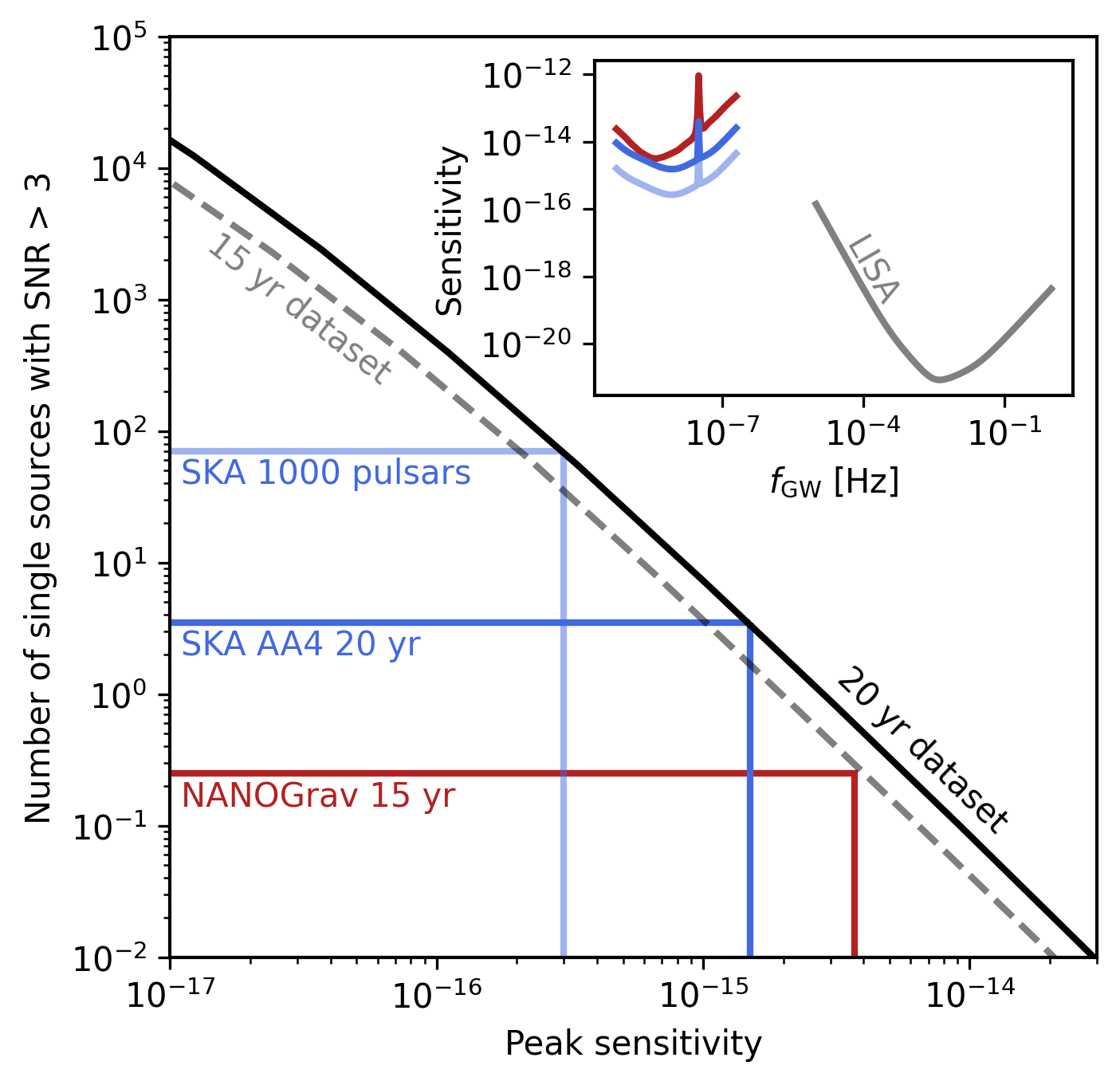}
    \caption{The number of expected single-source detections with SNR > 3 for a PTA sensitivity curve with a given peak sensitivity, for a 15~yr dataset or a 20~yr  dataset. Reference values are the current NANOGrav 15~yr reported sensitivity (in red, see \citealt{Agazie_et_al_2023_Detector}) and the projected SKAO-era sensitivity (in dark blue, for the projected 20-yr AA4 case with 174 pulsars, and in light blue, for a case wherein 1000 pulsars are tracked). The number of detectable sources scales roughly quadratically with improvements in sensitivity, yielding an expectation value of several to hundreds of detectable single sources in the SKAO-era. The inset plot shows the corresponding sensitivity curves in the context of the GW frequency band.}
    \vspace{-0.4cm}
    \label{fig:N_vs_peak}
\end{figure}

Here we provide a simple estimate of how the number of detectable single sources will increase relative to the current sensitivity of PTAs. To do so, two main ingredients are needed: (i) a model for the anticipated improvements in detector sensitivity, and (ii) a model for the astrophysical population of MBHBs. To establish a baseline sensitivity, we adopt the NANOGrav 15yr \citep[][]{Agazie_et_al_2023_Detector} sensitivity curve (shown in the inset of Figure~\ref{fig:N_vs_peak} in terms of dimensionless strain). To model the prospective sensitivity of a PTA in the SKAO (AA4) era, we instead use the sensitivity curve by \citet{Shannon_et_al_2025}, which assumes a 20-yr observation baseline and the tracking of 174 pulsars.

We define three scenarios:

\begin{itemize}

\item NANOGrav -- with 67 pulsars and 15~yr of total observation time.

\item SKAO (AA4) -- with 174 pulsars and 20~yr of total observation time \citep[][]{Shannon_et_al_2025}.

\item SKAO 1000 pulsars -- wherein $\mathcal{O}(10^3)$ pulsars are tracked \citep[][]{Keane_et_al_2025,Keane01.2026.SKA} for 20~yr.

\end{itemize}

For the latter scenario, we model the effect of increasing the number of pulsars, by simply improving the overall sensitivity of \citet{Shannon_et_al_2025} as $\propto N_{\rm p}^{-1}$, where $N_{\rm p}$ is the number of pulsars, following the scaling discussed by \mbox{\citet{2019Hazboun}}.

To model MBHB populations, we adopt a prescription based on the halo merger rates measured in the Millennium simulation \citep[][]{Fakhouri_et_al_2010}, which we explain in more detail below. It is important to emphasize that our goal here is not to achieve the most precise astrophysical prediction. Instead, we aim to illustrate the difference between current expectations for single-source detections and what may become feasible with the advent of SKAO. We note, however, that estimates based on galaxy merger rates gathered from more sophisticated models (e.g. the \textsc{holodeck} code; \citealt{Agazie_et_al_2023_Constraints,Kelley_et_al_2024}) are possible: recently, \citet{Shannon_et_al_2025} obtained estimates using a version of the \textsc{L-Galaxies} semi-analytical model \citep[][]{Izquierdo-Villalba_et_al_2022,Truant_et_al_2026}.

To construct populations of MBHBs, we begin with the differential halo merger rate (i.e. the number of mergers per halo per unit redshift per unit mass ratio),\\
\begin{align}
    \frac{{\rm d}^2 \Gamma}{{\rm d}\xi {\rm d}z} = B_1 \left( \frac{M_{\rm{halo}}}{10^{12}\,{\rm M}_{\odot}} \right)^{b_1} \xi ^{b_2} \exp \left[\left( \frac{\xi}{B_2} \right)^{b_3}\right] (1+z)^{b_4}\,,
    \label{eq:millennium}
\end{align}

\noindent where $\Gamma$ is the total number of mergers experienced by a halo of mass $M_{\rm halo}$ over the lifetime of the Universe. The variable $\xi \leq 1$ denotes the halo mass ratio in each merger, and the numerical parameters are given by $(B_1, B_2, b_1, b_2, b_3, b_4) = (0.0104, 9.72 \times 10^{-3}$, 0.133, -1.995, 0.263, 0.0993) \citep[][]{Fakhouri_et_al_2010}.

The corresponding MBH merger rate, $\dot{N}_{\bullet \bullet}$, is then obtained by coupling the halo merger rate with the BH mass function, ${\rm d}n_{\bullet}/{\rm d}M_{\bullet}$:\\
\begin{align}
    \frac{{\rm d} ^{3} \dot{N}_{\bullet \bullet}}{{\rm d}M_{\bullet} {\rm d}\xi {\rm d}z} &= {P_{\rm occ}}(M_{\rm{halo}},z) \frac{4 \pi c D^2_{\rm{com}}(z)}{(1+z)^3} \frac{{\rm d}n_{\bullet}}{{\rm d}M_{\bullet}}(M_{\bullet},z) \frac{{\rm d}^2 \Gamma}{{\rm d}\xi {\rm d}z}(\xi,z_{\rm{del}})\,,
    \label{eq:generalSMBBHrate}
\end{align}

\noindent where $D_{\rm com}(z)$ is the comoving distance at redshift $z$, $P_{\rm occ}$ is the fraction of halos that host MBHs, and the delayed redshift $z_{\rm del}$ accounts for the time lag between a halo merger and the eventual MBHB merger. For the MBH mass function, we adopt the results of \citet{Shankar_et_al_2013}, which are interpolated for redshifts between $0$ and $4$ \citep[see][for further details]{Zwick_et_al_2022,2024Stegmann}.

Finally, we use it together with the relation between halo mass and MBH mass proposed by \citet{Croton_2009}:\\
\begin{align}
    M_{\bullet}=\left[\frac{M_{\rm{halo}}(1+z)}{2\times10^7 \, {\rm M}_\odot}\right]^{3/2} \, {\rm M}_\odot.
\end{align}

With these ingredients, we can estimate the number of individually detectable MBHBs by integrating the differential merger rate. As a reference, we adopt an SNR threshold of three, following the convention in \citet{Shannon_et_al_2025} (though the exact threshold may vary depending on the source properties). The expected number of detections is then\\
\begin{align}
    N_{\bullet \bullet}^{\rm{det}} = \int \int \int \int \frac{{\rm d}^{3} \dot{N}_{\bullet \bullet}}{{\rm d} M_{\bullet} {\rm d} \xi {\rm d} z} \frac{1}{\dot{f}_{\rm{GW}}} \Theta \left({\rm{SNR}-3}\right)\,{\rm d} f_{\rm{GW}}\, {\rm d} \xi \, {\rm d} M_{\bullet} \, {\rm d} z\,,
    \label{Eq:Ndet}
\end{align}

\noindent where $\Theta$ is the Heaviside function. Here we have distributed binaries across frequency bins according to their GW residence time \citep{1964Peters}, $\dot{f}_{\rm GW}^{-1}$, where $f_{\rm GW}$ is the observed GW frequency. This is a lower limit, as environmental effects can increase the number of binaries that reside at relatively high frequency \citep{2017Kelley}.

For simplicity, we assume $P_{\rm occ} = 1$, i.e. every halo hosts an MBH, and we neglect any delay between halo mergers and MBHB mergers (i.e. $z_{\rm del} = z$). In this reference case, the SNR of each monochromatic source is given directly by its characteristic GW strain $h_{\rm c}$ relative to the detector sensitivity curve. Examples of such sensitivity curves are shown in the inset of Figure~\ref{fig:N_vs_peak}, while the characteristic strain can be computed with standard, sky-averaged formulas \citep[see, e.g.][]{2018gwv..book.....M}. Note that, in reality, the sky sensitivity of a PTA varies significantly with sky angle, due to the limited coverage of tracked pulsars. Here we are assuming isotropic sensitivity for the sake of simplicity and a fair comparison between array configurations, noting that our population estimate may have to be reduced by a factor of order unity considering the effective sky coverage.

The results of our population estimates are shown in Figure~\ref{fig:N_vs_peak}. We plot the expected number of detectable MBHBs as a function of the peak sensitivity of a PTA sensitivity curve for an observation time of 15 yr or 20 yr, highlighting the three scenarios mentioned above. The number of expected single sources scales approximately quadratically with improvements in sensitivity. For the NANOGrav 15~yr curve's peak sensitivity of $\sim$$3 \times 10^{-15}$ in terms of GW strain, only a few times $10^{-1}$ sources are expected. As the sensitivity approaches projected SKAO AA4 levels, this number increases to several high-SNR detections. Order $10^2$ single-source detections are projected if SKAO is able to track $\sim$$10^3$ pulsars. Despite the simplicity of our analysis, the emerging picture strongly suggests that PTAs in the SKAO era can expect to detect a substantial amount of high-SNR single sources in the nHz frequency band. When comparing our analysis with that performed in \citet{Shannon_et_al_2025}, based on the \textsc{L-Galaxies} merger rate, we note that their results are more optimistic, showing 30--40 single source detections in the SKAO AA4 configuration. However, we also remark that their analysis shows that NANOGrav 15 yr should have been able to see a few.

\subsection{Gas-embedded sources as multi-frequency gravitational-wave emittors}

MBHBs are expected to form following the major merger of galaxies \citep[][]{Begelman_et_al_1980}. Many of the individual sources that we expect to detect with SKAO-era PTAs are likely to be such binaries, the coalescence of which is driven by the dynamical and hydrodynamical processes triggered during the merger \citep[see, e.g.][]{Amaro-Seoane_et_al_2023}. In particular, it is well know that major galaxy mergers trigger large gas inflows towards the central regions of the remnant through mechanisms such as shocks and angular momentum transfer (see \citealt{Capelo_et_al_2023} for a recent review). Beyond inducing starburst episodes, this inflowing gas can assemble into a fully fledged accretion disc powering quasar activity.

The subsequent evolution of MBHBs embedded in gaseous environments is highly complex, and subject of innumerable analytical and numerical work \citep[e.g.][]{2002armitage,2021dan,Derdzinski_et_al_2021,2024siwek}. Interactions between the binary and the surrounding disc result in gaseous torques, which alter the evolution of its orbital elements and modify the vacuum-only, GW-driven inspiral rate. Moreover, the presence of gas can also lead to observable effects even at small binary separations, where GW back-reaction is fully dominant. These effects include, for example, dephasing of GW signals relative to the vacuum expectation, a topic of growing interest in view of the forthcoming LISA mission \citep[e.g.][]{1993chakrabarti,bara,2022garg,2023MNRAS.521.4645Z,2024garg}.

A central aspect of MBH-gas interactions is the distinction between secular torques and torque variability. The time-averaged (secular) torque determines the long-term orbital evolution of wide MBHBs and can additionally set the magnitude of GW dephasing during the GW-back-reaction-dominated inspiral. However, simulations ubiquitously show that short-term torque variability vastly dominates over the magnitude of the time-averaged torque, often by factors of hundreds. The origin of this strong variability is ultimately the non-linearity of the gas flow around the binary, as well as the several intrinsic sources of fluctuations and turbulence in accretion discs \citep[e.g.][]{2012roedig}, such as the magnetorotational instability (MRI).

Recent work by \citet{Zwick_et_al_2022} investigated the impact of arbitrary spectra of stochastic torques on MBHBs in the GW-driven regime \citep[see also][]{2025copparoni}. They showed that such torque fluctuations perturb the orbital velocity and separation of the binary on dynamical time-scales or faster, leading to the generation of additional harmonics in the GW signal, dubbed DWs. Crucially, this additional GW emission occurs at all frequencies represented in the torque spectrum, and is therefore not limited to the orbital frequency of the binary.

{\it This raises the intriguing prospect of detecting the same binary in two distinct frequency bands: the fundamental GW emission in the nHz range, and the additional harmonics from stochastic torques in the mHz range.} Here we present an estimate of such possibility, in light of the prospect of detecting many single sources with an SKAO-era PTA.\\

\subsection{Multi-band detection of gravitational waves}

Following the approach of \citet{Zwick_et_al_2022}, we model all sources of variability in the interaction between MBHBs and their gaseous environment by assuming that torque fluctuations possess a characteristic amplitude and follow a cascade-like spectrum. This analogy to turbulence reflects the idea that energy injected into the disc on large scales is transferred through a hierarchy of eddies down to smaller scales, with power distributed across frequencies according to a power law. Guided by this picture, we assume that the torque variability scales as\\
\begin{equation}
    \dot{L}(f_{\rm turb}) \propto f_{\rm turb}^{-n/2}\,,
\end{equation}

\noindent where $f_{\rm turb}$ is the frequency of the turbulence, i.e. the inverse of the eddie's size divided by the turbulent velocity, and $n$ parameterizes the slope of the turbulent cascade in energy. The choice of the value of $n$ depends on the precise physical mechanism driving the cascade, and for incompressible fluids we have $n = 5/3$ \citep[][]{Kolmogorov_1941}. In \citet{Zwick_et_al_2022}, it was shown that the relevant parameter for DW emission is, in fact, half of the power-law index that describes the cascade in energy, since perturbations to the binary parameters scale as the square root of the fluctuations in energy. Here, we choose for simplicity a reference value of $n=5/3$, though we note that many values between $3/2$ and $2$  have been shown to model turbulent-cascade mechanisms such as MRI in compressible accretion discs.

The remaining parameter required to describe the torque spectrum is a characteristic amplitude. Constraints on the strength of both secular torques and torque fluctuations for gas-embedded MBHBs have been obtained from numerical simulations. We focus here on binaries with a mass ratio of order $\sim$1, for which studies indicate that the secular hardening rate satisfies \citep[see, e.g.][]{2002armitage,2021dan,2024siwek}\\
\begin{equation}
    \frac{\dot{a}}{a} = \frac{\dot{E}}{E} \sim 3 \times \frac{\dot{M}}{M}\,,
\end{equation}

\noindent where $a$ is the semi-major axis, and $M$ and $E$ are the total binary mass and energy, respectively. The corresponding secular gas torques are well modelled by assuming the scaling of viscous torques \citep{2002armitage}:\\
\begin{equation}
    \dot{L}  \simeq 3 \times \dot{M} a^2 \Omega_{\rm K}\,.
\end{equation}

The ratio of the fluctuation amplitude at the orbital frequency $\Omega_{\rm K}$ to this secular torque has also been measured in simulations \citep{2012roedig}. For instance, in Mach 10 discs with equal-mass binaries, the ratio reaches values of order $\sim$30 \citep{2024chris}. This ratio is seen to grow significantly for thinner discs and for unequal-mass systems, with values up to $\sim$300 reported in intermediate-mass ratio inspiral simulations \citep{Derdzinski_et_al_2021}. Motivated by these results, we adopt a baseline model for torque variability similar to that of \citet{Zwick_et_al_2022}:\footnote{In \cite{Zwick_et_al_2022}, the torque model employed both linear Type I torques and viscous torques, depending on the mass ratio. Here we only use viscous torques, as they are appropriate for the mass ratios of interest.}\\
\begin{equation}
    \dot{L}_{\rm var}(f_{\rm turb}) \equiv 100 \times f_{\rm Edd}\,\dot{M}_{\rm Edd}\, a^2 \Omega_{\rm K} \, \left(\frac{f_{\rm turb}}{\Omega_{\rm K}}\right)^{-n}\,.
    \label{eq:stochtorq}
\end{equation}

Once again, the variable $f_{\rm turb}$ here represents the frequencies at which the torque variability is acting, and the torque spectrum is scaled to have values of $\sim$100 of the secular torque, as is expected from extrapolating current simulations for equal-mass binaries embedded in circumbinary discs to more realistic Mach numbers \citep[see, e.g. the discussion in][]{2024zwick}. We note that the strength of the DWs depends linearly on this typical fluctuation scale, and the associated DW background scales in quadrature.

With the torque spectrum fully characterised, we can proceed to compute the resulting perturbations to the binary's GW emission. As mentioned above, \citet{Zwick_et_al_2022} demonstrated that stochastic torques influence the GW signal of a gas-embedded binary by perturbing its quadrupole moment on short time-scales. As a result, nHz binaries subjected to such torques emit not only the standard quasi-monochromatic GW signal but also a spectrum of additional harmonics, the DWs:\\
\begin{align}
    h_{\rm DW}(f_{\rm GW}) \approx \frac{4G\dot{L}_{\rm var}(f_{\rm GW})}{c^4d_{\rm L}}\,,
    \label{eq:DW}
\end{align}

\noindent where GWs are produced at all $f_{\rm GW}= f_{\rm turb}$. To re-iterate, here $h_{\rm DW}$ is the time-domain GW strain that is produced in addition to the vacuum MBHB emission, due to the presence of high-frequency orbital perturbations caused by torque variability. A key feature of this additional GW emission is that it can extend to frequencies higher than the main binary orbital frequency, reaching the sensitivity band of LISA.

\begin{figure}
    \centering
    \includegraphics[width=0.49\linewidth]{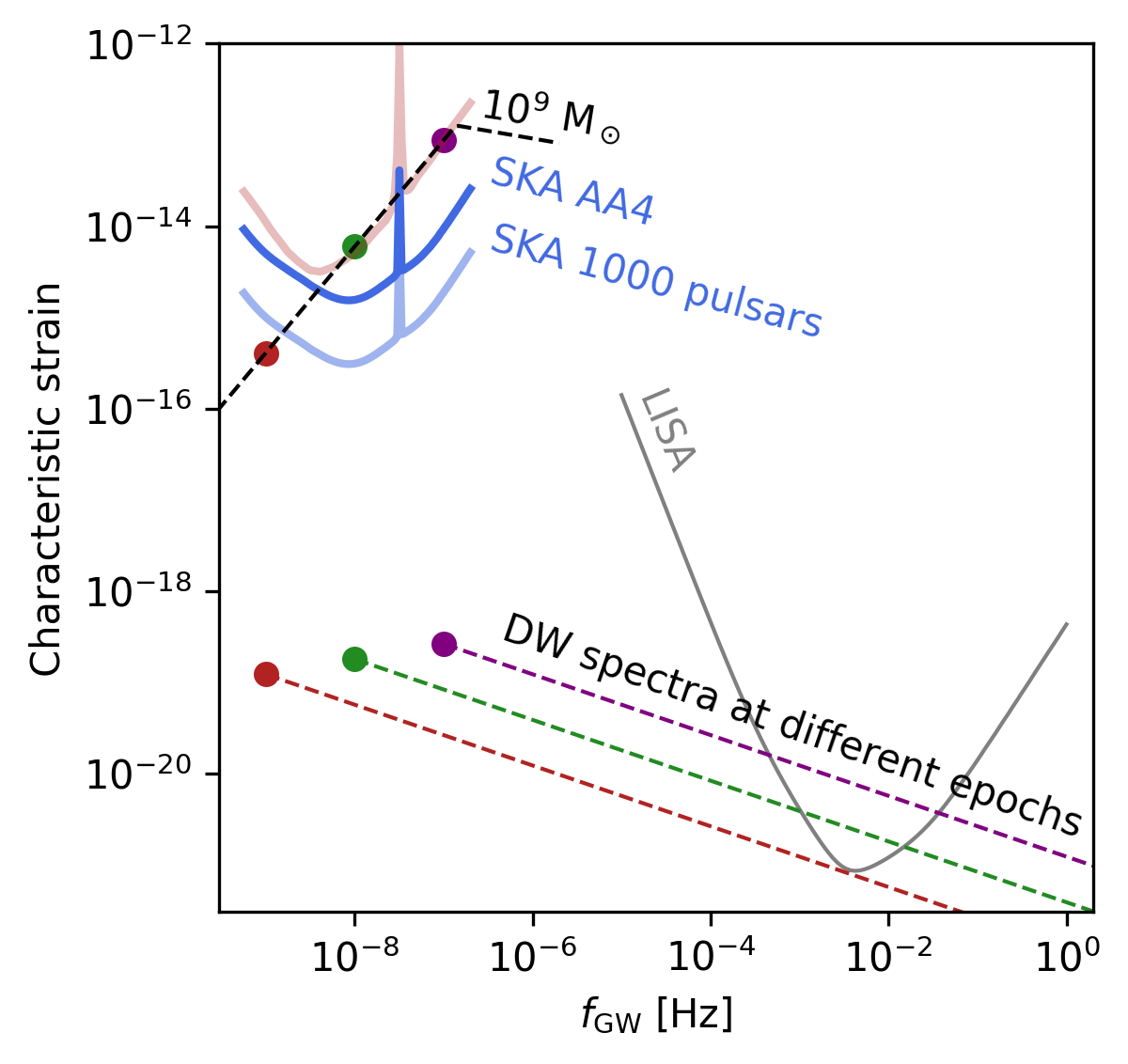} \includegraphics[width=0.50\linewidth]{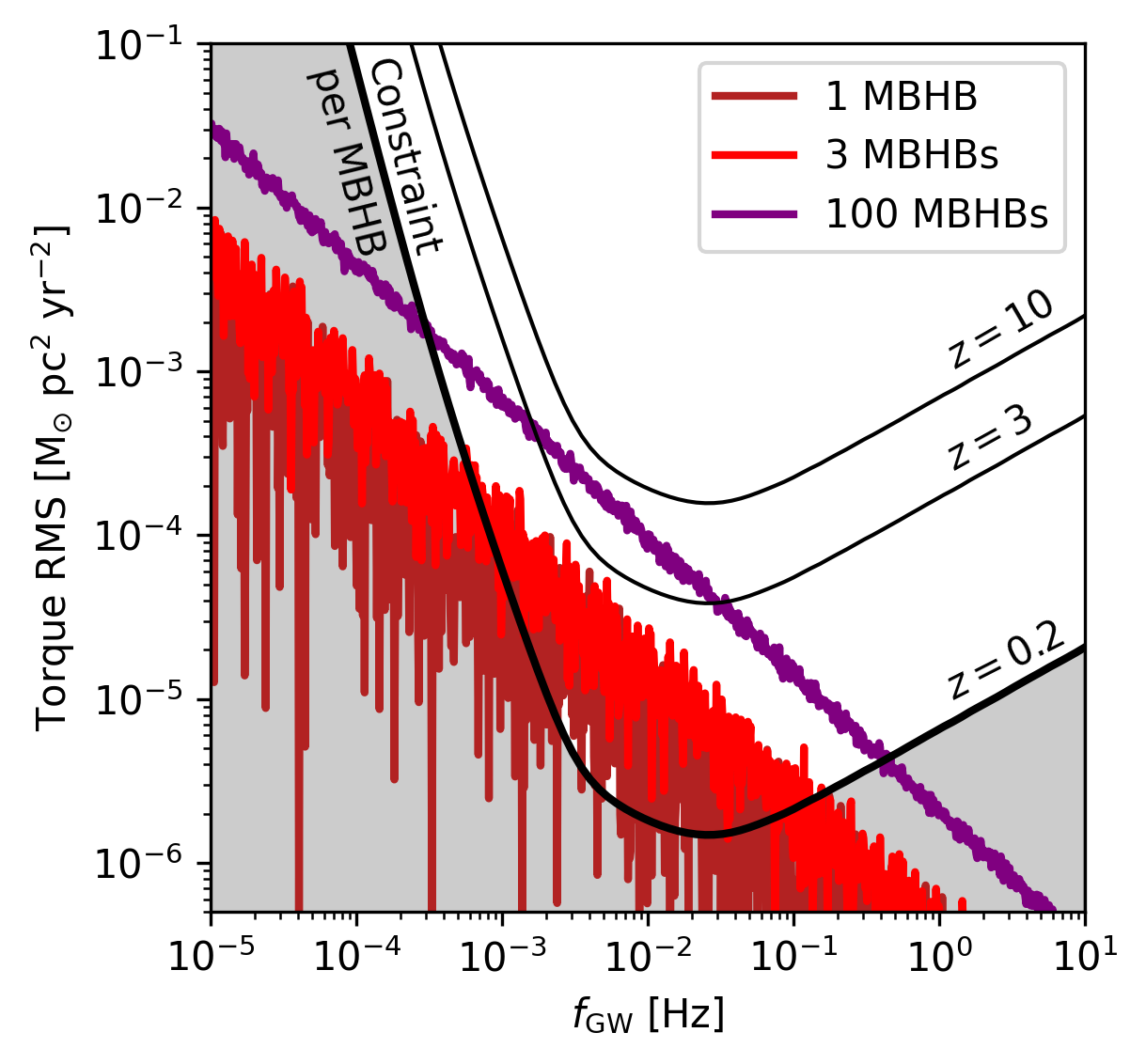}
    \caption{Left-hand panel: we show the characteristic strain evolution of a gas-embedded $10^9 \,{\rm M_\odot}$ equal-mass MBHB at $z = 0.2$. The dashed black line traces the binary’s inspiral, with markers at $f_{\rm GW}=10^{-9}$, $10^{-8}$, and $10^{-7}$ Hz. The dashed coloured curves show additional GW harmonics induced by stochastic torques from a circumbinary disc accreting at $f_{\rm Edd}=1$. Right-hand panel: LISA 4-yr sensitivity to DWs, shown in black as the minimum RMS torque fluctuation amplitude required for detectability at different redshifts. The brown, red, and purple spectra illustrate an increasing amount of DW sources with the same parameters as in the left-hand panel (1, 3, and 100, respectively) adding in quadrature. The system parameters here are simply chosen for illustration purposes, but represent a plausible source.}
    \label{fig:DW}
\end{figure}

The left-hand panel of Figure~\ref{fig:DW} illustrates a realization of a system capable of emitting GWs observable both in the PTA and LISA bands. We plot the characteristic strain track of an equal-mass binary with a total mass of $10^9$ ${\rm M_\odot}$ located at $z = 0.2$, highlighting three representative stages of its evolution (at $f_{\rm GW} = 10^{-9}$, $10^{-8}$, and $10^{-7}$~Hz). Assuming that the binary is embedded in a circumbinary disc and accreting at the Eddington rate ($f_{\rm Edd} = 1$), we employ Equation~\eqref{eq:stochtorq} to estimate the spectrum of stochastic torques acting on the system. Using Equation~\eqref{eq:DW}, we then compute the resulting spectrum of additional GW harmonics induced by these torques. The corresponding characteristic strain is obtained by multiplying Equation~\eqref{eq:DW} by $(T_{\rm LISA} \, f_{\rm GW})^{1/2}$, where $T_{\rm LISA} = 4$ yr represents the total observation time for LISA in which DW power can be accumulated. Note that, over this time, the frequency of the MBHB will only drift by a small fractional amount. While important for signal extraction, this drift will not significantly change the interaction between the MBHB and its accretion disc, and therefore our results. Whether the resulting DW spectrum is detectable depends on several factors, including the binary’s intrinsic parameters (particularly, its mass and distance), the Eddington ratio, and the slope of the turbulent power spectrum. Nevertheless, with our representative choices discussed above, we find that especially massive and nearby MBHBs can indeed emit detectable GW signals in both the nHz and mHz frequency bands (and potentially in between, thus detectable by proposed $\mu$Hz detectors, such as $\mu$Ares and Doppler tracking; \citealt{Sesana_et_al_2021,Zwick_et_al_2025}; see also \citealt{Jenkins01.2026.SKA} for a complementary approach using the SKAO). While rare, this constitutes a truly unprecedented class of multi-band sources, that can only be accessed by using both LISA and PTAs in the SKAO era. Its detection becomes a realistic possibility in the case that PTAs are able to detect numerous single sources.

The right-hand panel of Figure~\ref{fig:DW} shows the typical constraints that 4~yr of LISA observations can place on DWs emitted at different redshifts. The constraint is expressed in terms of the required root-mean-square (RMS) amplitude of torque fluctuations at a given frequency $f_{\rm turb}$, necessary to generate DWs with sufficient power. This provides a direct means to test whether the results of hydrodynamical simulations indicate that DW will be relevant. We also illustrate an additional aspect discussed in \citet{Zwick_et_al_2022}, namely that DWs from multiple binaries can add in quadrature, forming a stochastic DW background. Here, we simply add the contributions from three and hundred identical sources, though in reality this will arise from a varied population of massive BH binaries.

\subsection{Conclusions}

In the SKAO era, PTAs are expected to confidently detect several massive MBHBs as individual GW sources. Many of these systems are embedded within gaseous circumbinary discs, where their evolution is shaped by a combination of secular and stochastic torques. While the secular component governs the long-term orbital evolution, stochastic torques can modulate the instantaneous orbital dynamics, thereby altering the GW emission. As a result, the emitted GWs form a spectrum of additional harmonics, that mirror the spectrum of the stochastic perturbations. We have shown that, in some cases, particularly for massive, nearby binaries, this spectrum of DWs can extend into the mHz regime, producing signals potentially detectable by LISA.

DWs can manifest either as discrete contributions or as part of a stochastic background. Quantifying the amplitude of this background and evaluating the prospects of resolving its individual components will require more sophisticated models of both the binary population and the coupling between orbital time-scales and the turbulent cascade within the surrounding gas. Additionally, it will be necessary to develop strategies to match the nHz GW background produced by SMBH binaries with the mHz DW background observed in LISA, to confirm that they originate from the same population of sources. For example, this could be achieved by exploiting some form of correlation (e.g. in the GW phase or in the anisotropy) between the nHz GW background and the mHz DW background. Whether these issues can be fully resolved requires further study.

However, even with simplified estimates, our analysis underscores the exciting possibility of \textit{simultaneous GW detections in both the PTA and LISA bands from the same sources}. At a minimum, such multi-band detections would offer valuable insights into the demographics of gas-embedded MBHBs and the physics of turbulence and stochastic processes in circumbinary discs. Conversely, if the DW background proves to be sufficiently strong, it could even interfere with certain LISA science objectives by polluting the detector's sensitivity band. This highlights the need to understand the properties of nHz binaries and their DW emission in detail. Future single-source detections in the SKAO era could therefore not only reveal nearby MBHBs but also provide essential priors for mitigating DW contamination in the LISA data.\\\\\\

{\bf Acknowledgements}\\
We acknowledge useful correspondence with Philipp Denzel, Massimo Dotti, Kayhan G\"ultekin, David Izquierdo-Villalba, and Mark Sargent. PRC acknowledges support from the Swiss National Science Foundation under the Sinergia Grant CRSII5\_213497 (GW-Learn). LZ is supported by the European Union’s Horizon 2024 research and innovation program under the Marie Sk\l{}odowska-Curie grant agreement No. 101208914. The Center of Gravity is a Center of Excellence funded by the Danish National Research Foundation under grant No. 184.\\

\bibliographystyle{abbrvnat-maxbibnames4}
\bibliography{SKA-LISA}

\end{document}